\documentclass[preprintnumbers, pra, showpacs, floatfix,twocolumn,
preprintnumbers, letterpaper, superscriptaddress]{revtex4-2}
\usepackage{float}
\usepackage{amsfonts}
\usepackage{amsmath}
\usepackage{amssymb,epsf}
\usepackage{latexsym}
\usepackage{graphicx,epsfig}
\usepackage{amssymb}
\usepackage{subfigure}
\usepackage[colorlinks=true,citecolor=blue,linkcolor=blue,urlcolor=black]{hyperref}
\usepackage{epstopdf}
\usepackage{color}

\def\be{\begin{equation}}
\def\ee{\end{equation}}
\def\ba{\begin{eqnarray}}
\def\ea{\end{eqnarray}}
\def\la{\langle}
\def\ra{\rangle}

\begin{document}
\title{Conditional global entanglement in a Kosterlitz-Thouless quantum phase transition }
\author{Elahe Samimi}
\email{elh.samimi@gmail.com}
\affiliation{Department of Physics, School of Science, Shiraz University, Shiraz 71946-84795, Iran}

\author{Mohammad Hossein Zarei}
\email{mzarei92@shirazu.ac.ir}
\affiliation{Department of Physics, School of Science, Shiraz University, Shiraz 71946-84795, Iran}

\author{Afshin Montakhab}
\email{motomonty@gmail.com}
\affiliation{Department of Physics, School of Science, Shiraz University, Shiraz 71946-84795, Iran}

\begin{abstract}
Entanglement is known as an important indicator for characterizing different types of quantum phase transitions (QPTs), however it faces some challenges in the Kosterlitz-Thouless (KT) phase transitions due to an essential singularity which cannot be identified in finite derivatives of the ground state energy. In this paper, we consider global entanglement (GE) in a KT phase transition and show that while it does not indicate any clear signature of the phase transition, the conditional version of GE is a good indicator with strong signatures of the KT transition. In particular, we study a deformed version of the $Z_d$ Kitaev model which has an intermediate KT phase which separates a $Z_d$ topological phase from a magnetized phase at two different KT transition points. Using a mapping to the classical $d$-state clock model, we consider GE and the generalized GE and show that they do not provide a reliable indicator of transition points. However, their difference called conditional global entanglement (Q) shows a peak at the first KT transition point. Additionally, we show that it can characterize various phases of the model as it behaves substantially different in each phase. We therefore conclude that Q is a useful measure that can characterize various phases of KT QPTs as well as their related critical points.
\end{abstract}
 \maketitle
\section{Introduction}\label{sec1}
Studying correlations in physical systems is one of the most important approaches for considering transitions between different phases of matter \cite{kardar2007statistical,goldenfeld2018lectures}. In particular, QPTs are qualitatively different in a sense that their critical fluctuations can be understood in the notion of quantum mechanics \cite{sachdev1999quantum,carr2010understanding} where entanglement  is a powerful tool. It specifically has attracted much attention in topological quantum phase transitions (TQPTs) which have been extensively studied \cite{zhang2022multipartite,hamma2008entanglement,chen2010quantum} thanks to the different nature of entanglement in topological phases \cite{levin2006detecting,wen2013topological}. The role of entanglement becomes more interesting when one studies KT phase transitions where there are many challenges due to the essential singularity which appears in the ground state energy \cite{li2020critical,zarei2020kosterlitz,justino2012bell,yang2007ground,zhang2021fidelity,sun2015fidelity}. What emerges out of these notions is that there is a need to choose a reasonable measure of entanglement for characterizing different types of phase transitions \cite{beckey2021computable,amico2008entanglement}. 

Among different measures of entanglement, multipartite measures are suitable candidates for probing QPTs. This is motivated by the fact that phase transitions are accompanied by the divergence of the correlation length at the critical point, which results in the appearance of long-range correlations at criticality. This fact leads to the expectation that multipartite entanglement reaches the maximum value at the critical point in QPTs \cite{bayat2017scaling,de2006genuine,rigolin2006operational,hauke2016measuring,de2006multipartite}. Among different multipartite measures, GE \cite{meyer2002global} has been well-known through various scientific studies \cite{radgohar2018global,montakhab2010multipartite,montakhab2008dynamics,de2006genuine,lakshminarayan2005multipartite,vimal2018quantum,rigolin2006operational,da2017many,cui2008multiparticle,samimi2022global,zarei2022foliated}. 
Although it peaks at the critical point in ordinary QPTs \cite{de2006genuine,rigolin2006operational}, it has been shown that it behaves monotonically through the TQPT \cite{samimi2022global}. While the same behavior has also been observed \cite{samimi2022global} for the generalized version of global entanglement ($\widetilde{ \text{GE}}$), the conditional global entanglement denoted by Q shows the desired behavior by peaking at the critical point of a TQPT \cite{samimi2022global}. This result suggests that Q, as a measure of multipartite entanglement, should be considered for studying other types of QPTs.

On the other hand, calculating multipartite entanglement for different quantum systems needs costly computational methods such as tensor networks \cite{cirac2021matrix,shi2016geometric}. However, there are also mappings in spin models which establishes a quantum-classical correspondence \cite{castelnovo2008quantum,castelnovo2005quantum,zarei2019phase,zarei2022foliated,samimi2022global,zarei2018dual,zarei2020classical,zarei2019ising,zarei2021topological}. Regarding such mappings, there is a possibility to map entanglement measures to some classical quantities in the classical spin models. Since simulation of classical spin models is computationally less costly than quantum systems, one is able to do a simple analysis of entanglement in quantum systems, see \cite{samimi2022global}. For example, it has been shown that a deformation of the $Z_d$ Kitaev state denoted by $|K_d(\beta)\ra$, where $\beta$ is a control parameter, passes through KT QPTs where there is a mapping to the classical  d-state clock model \cite{zarei2020kosterlitz}. In this regard, $|K_d(\beta)\ra$ moves from the fully magnetized phase to the $Z_d$ topological phase with decreasing $\beta$. There is also an intermediate KT phase which appears between the trivial and the topological phase where transition points are of KT type. However, there is an important problem that, since singularity in such transitions is essential, it becomes a challenging task to find simple measures to characterize KT transitions. On the other hand, it is known that the correlation length shows divergence in the entire intermediate KT phase. Therefore, regarding the effect of the infinite correlation length on the multipartite entanglement, multipartite measures such as GE should be a good candidate for studying the aforementioned KT phase transitions.

In this regard, here we choose to analyze the behavior of GE and its conditional version in the above KT QPTs by using the mapping to the d-state clock model. Using such mapping, we simulate the d-state clock model to obtain GE, $\widetilde{ \text{GE}}$ and Q for $|K_d(\beta)\ra$. We find that both GE and $\widetilde{ \text{GE}}$ have monotonic behavior at the QPT. The derivatives of GE and $\widetilde{ \text{GE}}$ also fail to characterize KT QPTs, as their peaks do not appear near the critical points. However, Q develops a peak at the first critical point and decreases linearly within the KT phase. Near the second critical point, the behavior of Q versus $\beta$ modifies such that it converts to the power-law behavior. In other words, Q decreases linearly within the KT phase while in the topological phase the decay of Q becomes stronger (superlinear). This type of behavior can be related to the underlying behavior of correlation length in these two phases \cite{li2020critical}. Therefore, one can conclude that not only is Q able to indicate the KT QPT by peaking near the first critical point, but it also has a distinct behavior within the KT phase. Hence, it can be regarded as a reasonable measure for quantifying both TQPTs and KT QPTs.

The organization of the paper is as follows. In Sec. (\ref{sec2}), we define a $Z_d$ perturbed Kitaev model which exhibits an intermediate KT quantum phase. Moreover, we explain the quantum-classical correspondence between the above quantum model and the classical d-state clock model. In Sec. (\ref{sec3}), we show our numerical results on the behavior of GE and $\widetilde{ \text{GE}}$ as multipartite entanglement measures in the KT QPTs. In Sec. (\ref{sec4}), we present our numerical results for the behavior of Q in the aforementioned KT QPTs and show that it can capture such QPTs as well. We finally summarize our results and discuss some concluding remarks in Sec. (\ref{sec5}).

%For $d=2,3$ the clock model recovers to the classical Ising model and the Potts model, respectively. Note that for $d=4$, the model is the two copy of the Ising model. For $2 \leq d\leq  4$, as the temperature increases from $0$ to $\infty$ a second order phase transition separates the disordered  phase from the ordered phase at the critical temperature $T^{*}$. However, it has been known that for $d \geq 5$, the model has two transition temperatures where theoretical and numerical analysis predicts that both phase transitions are of BKT type \cite{ortiz2012dualities,li2020critical,li2020accurate,chen2022monte}. 
\section{The model}\label{sec2}
\begin{figure}[t]
\centering
\includegraphics[width=7cm,height=6cm,angle=0]{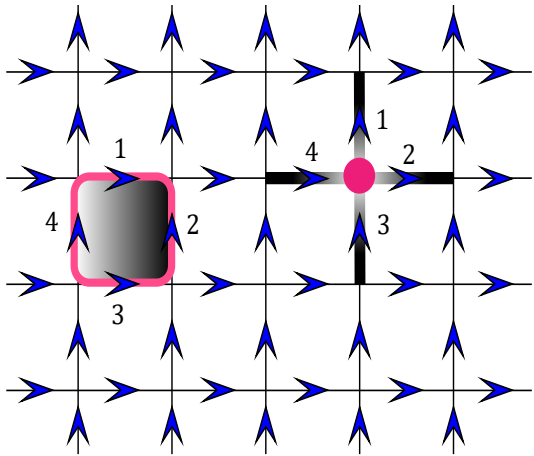}
\caption{(Color online) A directed square lattice with periodic boundary condition in 2D. Qudits exist on the edges. A plaquette and a vertex have been marked.} \label{fig:1}
\end{figure}

We start by reviewing the $Z_d$ Kitaev state which is a generalized version of the $Z_2$ Kitaev model where individuals are d-dimensional quantum systems called qudits. 
\begin{figure*}
	\centering     %%% not \center
	\subfigure[]{\label{fig:2a}\includegraphics[width=60mm, ,height=60mm]{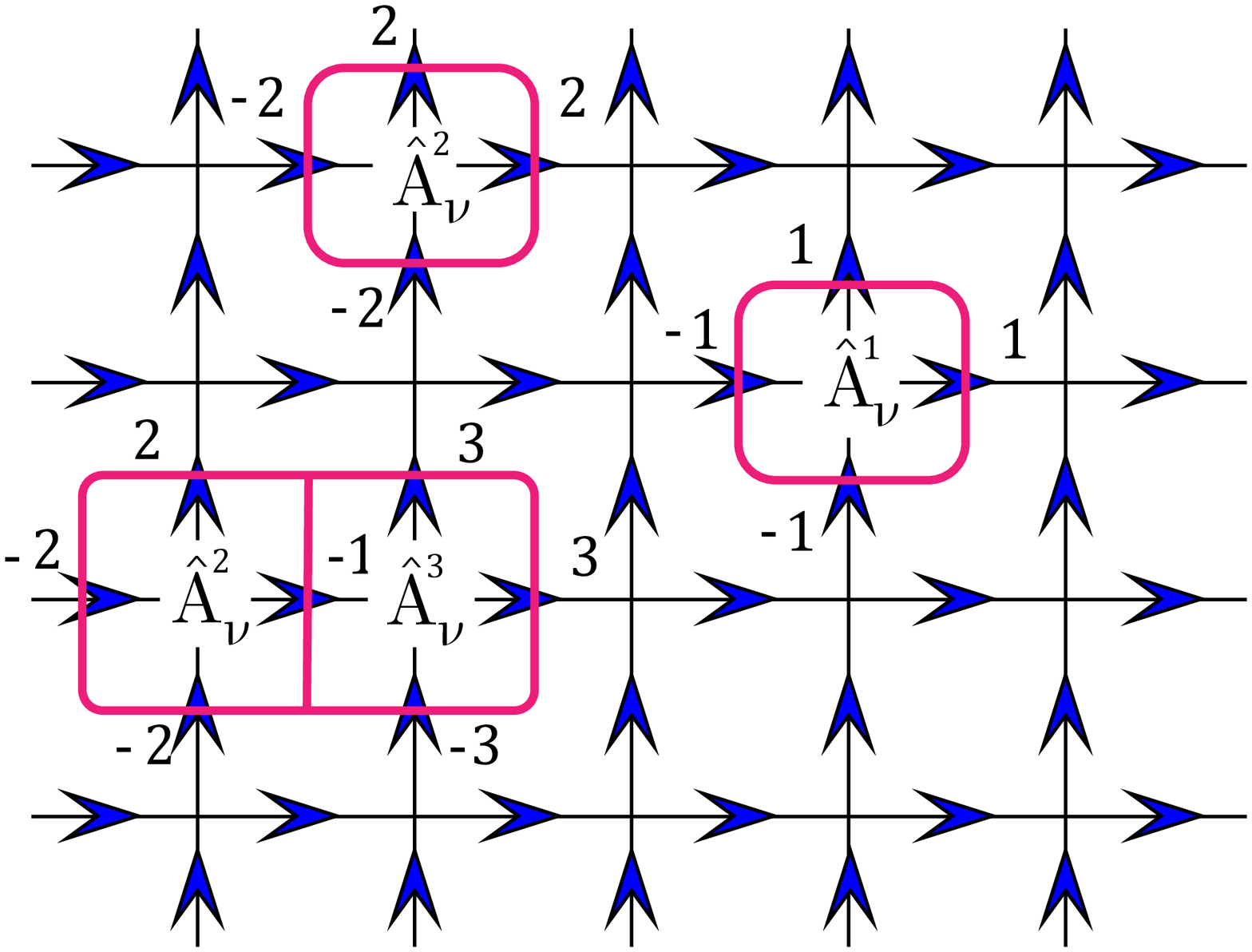}}
	\hspace{12mm}
	\subfigure[]{\label{fig:2b}\includegraphics[width=60mm, height=60mm]{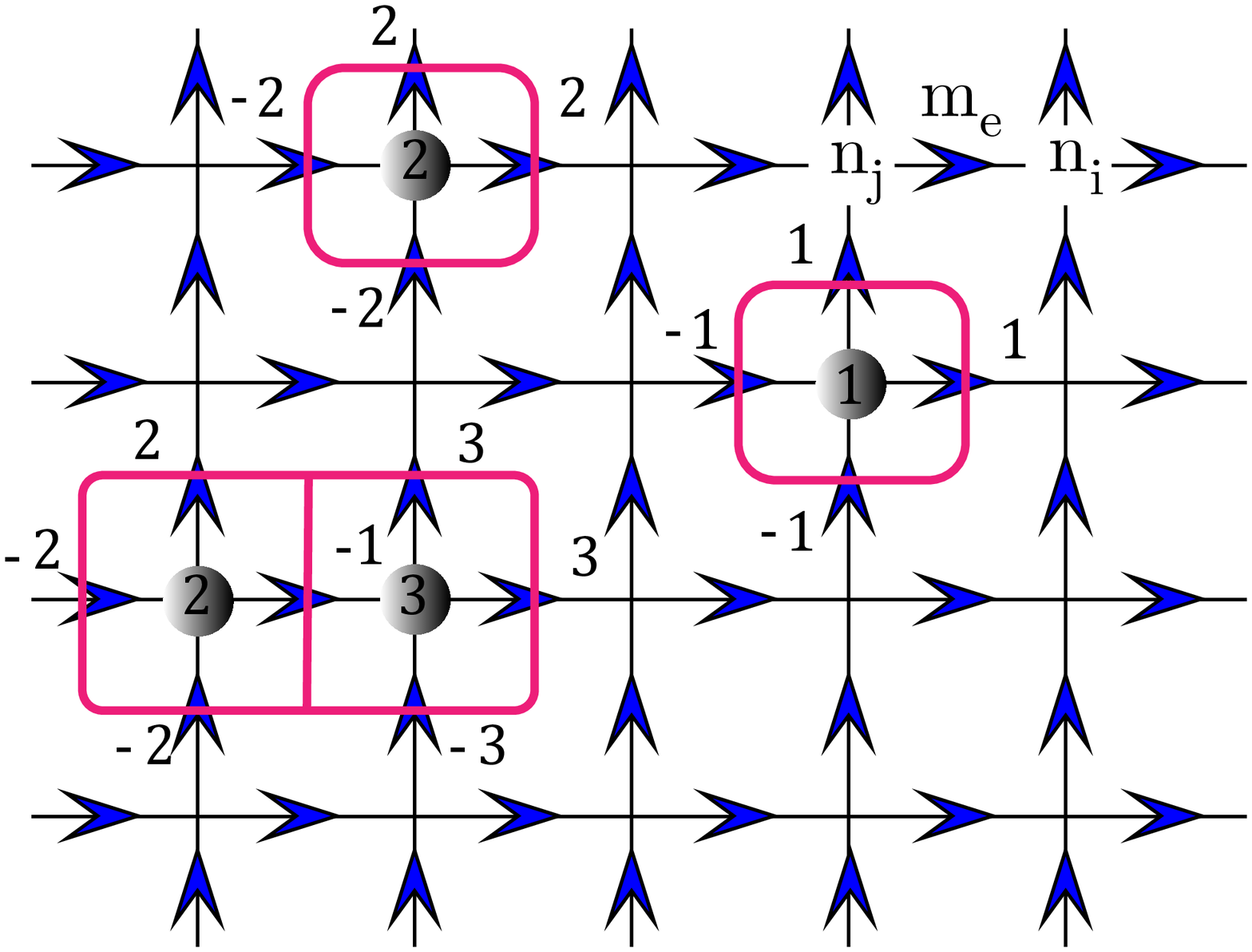}}
	\caption{(Color online) (a) Each $\hat{A_{v}}$ (vertex operator) corresponds to a loop in the dual lattice. The power of $\hat{A_{v}}$ determines the weight of the loop in the quantum model (b) d-state variables in the classical clock model are placed on the vertices. Circles mark variables with $ n_{i}\neq 0$. $m_{e}=n_{j}-n_{i}$ refers to edge variable. Each configuration in the clock model corresponds to a weighted loop pattern.}
\end{figure*}
Consider a 2D directed square lattice, see Fig. \ref{fig:1}, with periodic boundary condition where $N$ qudits are attached on the edges of the lattice. Then the $Z_d$ Kitaev state is a quantum stabilizer state stabilized by an Abelian group of stabilizer operators which are generated by generalized Pauli operators for a collection of $N$ qudits. In particular, the generalized version of the Pauli operators $\hat{Z}$ and $\hat{X}$ for a d-dimensional quantum system are defined in the following form:
\begin{equation}\label{pauli}
\hat{Z}= \sum_{m=0}^{d-1} w^{m}|m\ra\la m| ~~,~~\hat{X}=\sum_{m}|m+1\ra\la m| 
\end{equation}
where $w=\text{exp}(2\pi i/d)$ and $|m\ra$'s are the eigenvectors of $\hat{Z}$.
 Obviously, $\hat{Z}$ and $\hat{X}$ operators are unitary, but not Hermitian and they satisfy the relations $\hat{Z}^d=\hat{X}^d=\hat{I}$ and $\hat{Z}\hat{X}=w\hat{X}\hat{Z}$. Now it is possible to define two types of operators corresponding to plaquettes and vertices of the lattice. For each plaquette of the lattice, like the one marked in Fig. \ref{fig:1}, the plaquette operator $\hat B_{p}$ is defined in the following form:  
 
\begin{equation}\label{sta1}
\hat B_p =\hat{Z_1}\hat{Z}^{-1}_{2}\hat{Z}^{-1}_{3}\hat{Z_4},
\end{equation}
where the power of $\hat{Z}$ is $+1 (-1)$, if the corresponding edge of the plaquette has clockwise (counterclockwise) direction. Similarly, the star operator $\hat A_{v}$ corresponding to each vertex of the lattice, as shown in Fig. \ref{fig:1}, is defined by:
\begin{equation}\label{sta2}
\hat A_v =\hat{X_1}\hat{X_2}\hat{X}^{-1}_3\hat{X}^{-1}_4.
\end{equation}
where the power of $\hat{X}$ is $+1$ $(-1)$ if the link is outward (inward) to the vertex $v$. 
All plaquette and star operators commute, simply because they have even number of qudits in common. Therefore, they can construct the group of stabilizers. 
Let $|K_d\ra$ denotes the $Z_d$ Kitaev state, then it must be stabilized through all elements of the stabilizer group such that for each plaquette and star operator it satisfies $\hat{B}_p|K_d\ra=|K_d\ra$ and $\hat{A}_v|K_d\ra=|K_d\ra$. The resultant quantum state up to a normalization factor is:
\begin{equation}\label{kda1}
|K_d\ra =\prod_{v}(1+\hat{A}_v+\hat{A}_v^2+...+\hat{A}_v^{d-1})|0\ra^{\otimes N},
\end{equation}
where $|0\ra^{\otimes N}$ is a fully magnetized state.
Each $\hat A_{v}$ corresponds to a closed loop in the dual lattice. The power of $\hat A_{v}$ in Eq. \ref{kda1} determines the weight of the closed loop, see Fig. \ref{fig:2a}. This readily implies that $|K_{d}\ra$ is a string-net condensed state which can be represented as a superposition of all closed weighted loops.

Now Let us consider a deformation of the $Z_d$ Kitaev state such that it passes through a QPT. To this end, as it has been done in \cite{zarei2020kosterlitz}, we add fluctuations to the probability amplitudes of $|K_{d}\ra$ by applying a local invertible transformation to each qudit in the following form:
\begin{equation}\label{kdp}
|K_d(\beta)\ra \rightarrow \text{exp}\left\{ \frac{\beta}{2}\sum_{i}(\hat Z_{i}+\hat Z^{-1}_{i})\right\}|K_{d}\ra
\end{equation}
where $\beta$ is a control parameter and $i$ refers to the number of each qudit. In the $\beta \rightarrow \infty$ limit, each local invertible transformation $\text{exp}\left\{\frac{\beta}{2}(\hat{Z}+\hat{Z}^{-1})\right\}$ goes to a projection operator $|0\ra \la 0|$, simply because the probability amplitude of  $|0\ra \la 0|$ in the $\text{exp}\left\{\frac{\beta}{2}(\hat{Z}+\hat{Z}^{-1})\right\}=\sum_{m=0}^{d-1}\text{exp}\left\{ \beta \cos{\frac{2\pi m}{d}}\right\}|m\ra \la m|$ dominates. Therefore, $|K_{d}(\beta)\ra$ approaches the trivial state $|0\ra^{\otimes N}$. On the other hand, when $\beta = 0$ the $Z_d$ Kitaev state retrieves. Hence, it is obvious that a QPT must occur as a result of changing $\beta$ from $0$ to $\infty$. Moreover, note that we need also a normalization factor for $|K_d(\beta)\ra$ which is in the form of :
\begin{equation}\label{ui}
1/(\la K_d|\text{exp}\{\beta\sum_{i}(\hat{Z}_{i}+\hat{Z}_{i}^{-1})\}|K_d\ra)^{1/2}.
\end{equation}

To illustrate the QPT, let us explain an interesting property of $|K_{d}(\beta)\ra$. In particular, there is a mapping to the $d$-state clock model where the probability amplitudes of $|K_d(\beta)\ra$ correspond to the Boltzmann weights of different configurations of the d-state clock model in 2D which is described by the Hamiltonian:
\begin{equation}\label{clock}
H = -\sum_{\la ij \ra} \cos(\theta_{i}-\theta_{j}),
\end{equation}
where $\theta_{i}=2\pi n_{i} /d$, and $n_{i}=\{{0,1,...,d-1}\}$ refers to the classical d-state spins living in the vertices of a square lattice. In order to clarify the above quantum-classical correspondence, let us switch to the graphical representation of the clock model. Consider a 2D square lattice with periodic boundary condition and give each edge a direction as shown in Fig. \ref{fig:2b}. It is then possible to show an arbitrary configuration of spins in terms of closed weighted loops in the dual lattice. To this end, consider an arbitrary configuration shown in Fig. \ref{fig:2b} where for spins denoted by circles $n_{i}\ne0$ and for other spins $n_{i}=0$. Then, let $m_{e}=n_{j}-n_{i}$ denotes a different variable living in any edge of the lattice. By connecting lines which cross from edges with $m_{e}\ne 0$, a configuration with closed weighted loops emerges. It is perfectly obvious that the closed loops in the classical and quantum model are exactly the same.

In this regard, there is a relation between the partition function of the classical model and the normalization factor of the $|K_d (\beta)\ra$ \cite{zarei2020kosterlitz}. The partition function of the classical d-state clock model at temperature T is expressed as:
\begin{equation}\label{Zclock}
\mathcal{Z}_{clock}(T)=\sum_{\{\theta_{i}\}}\text{exp}\sum_{\la ij \ra}\cos(\theta_{i}-\theta_{j})/T.
\end{equation}
\begin{figure*}[t!]
	\centering     %%% not \center
	\subfigure[]{\label{fig:3a}\includegraphics[width=60mm, height=40mm]{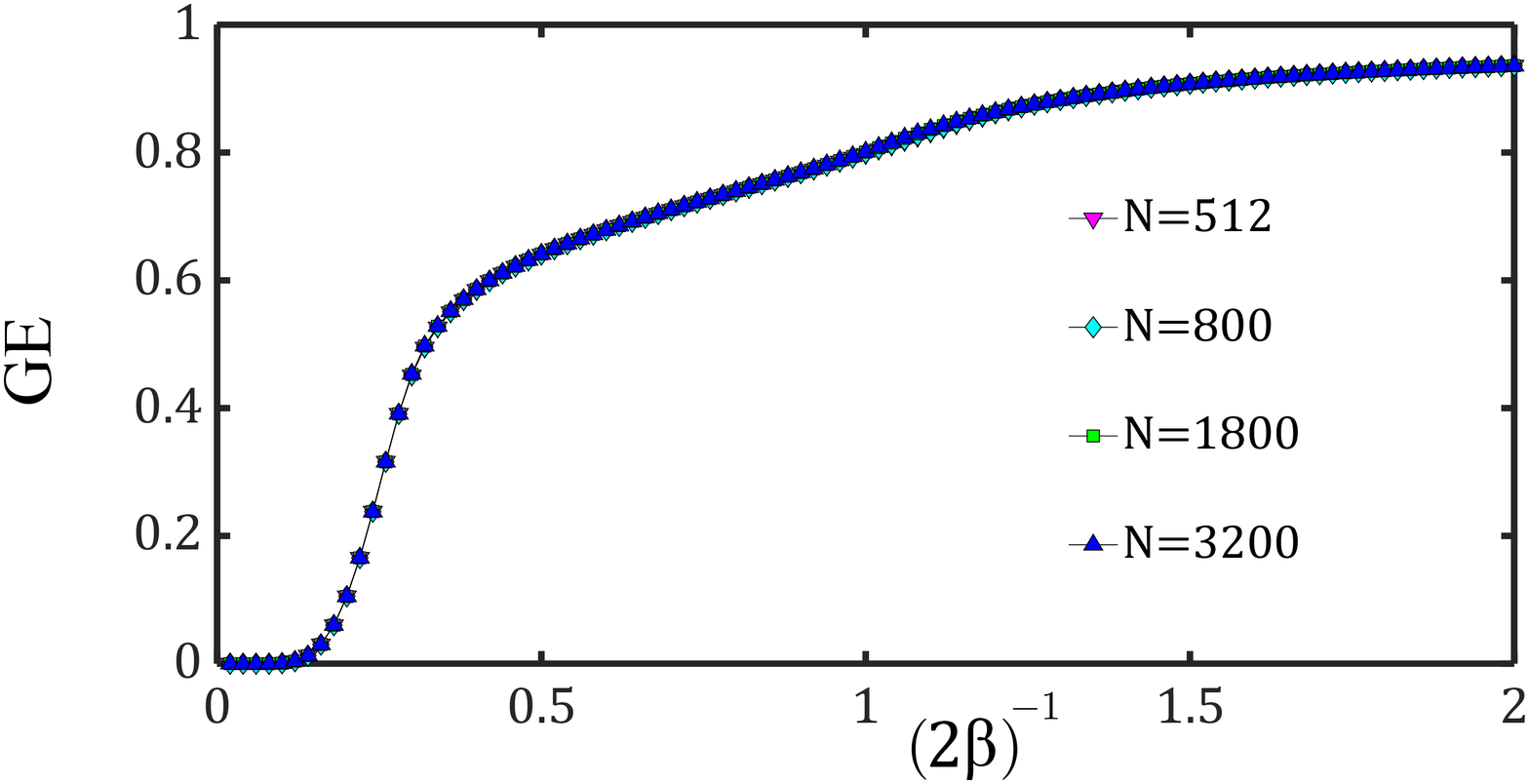}}
	\subfigure[]{\label{fig:3b}\includegraphics[width=60mm, height=40mm]{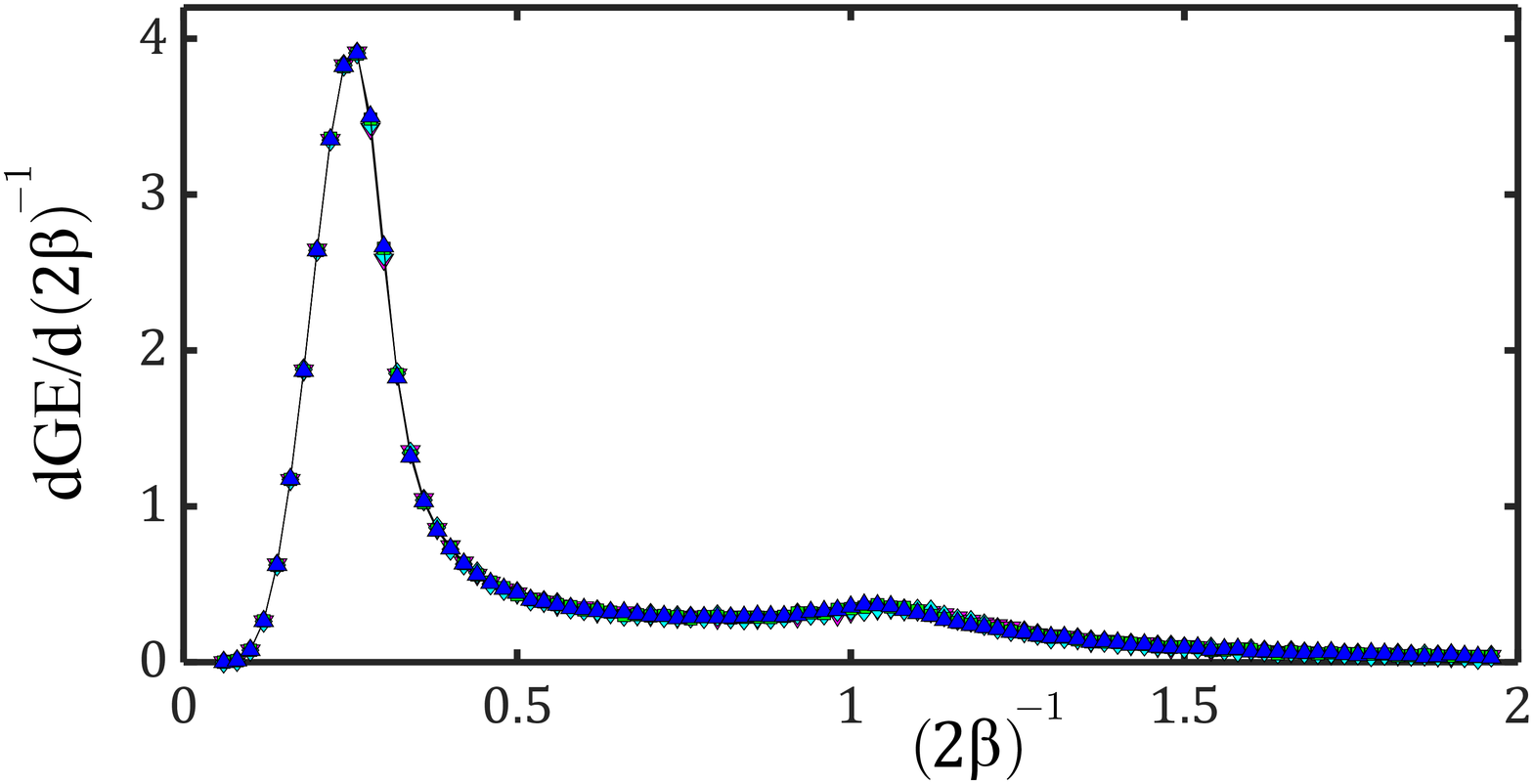}}
	\subfigure[]{\label{fig:3c}\includegraphics[width=60mm, height=40mm]{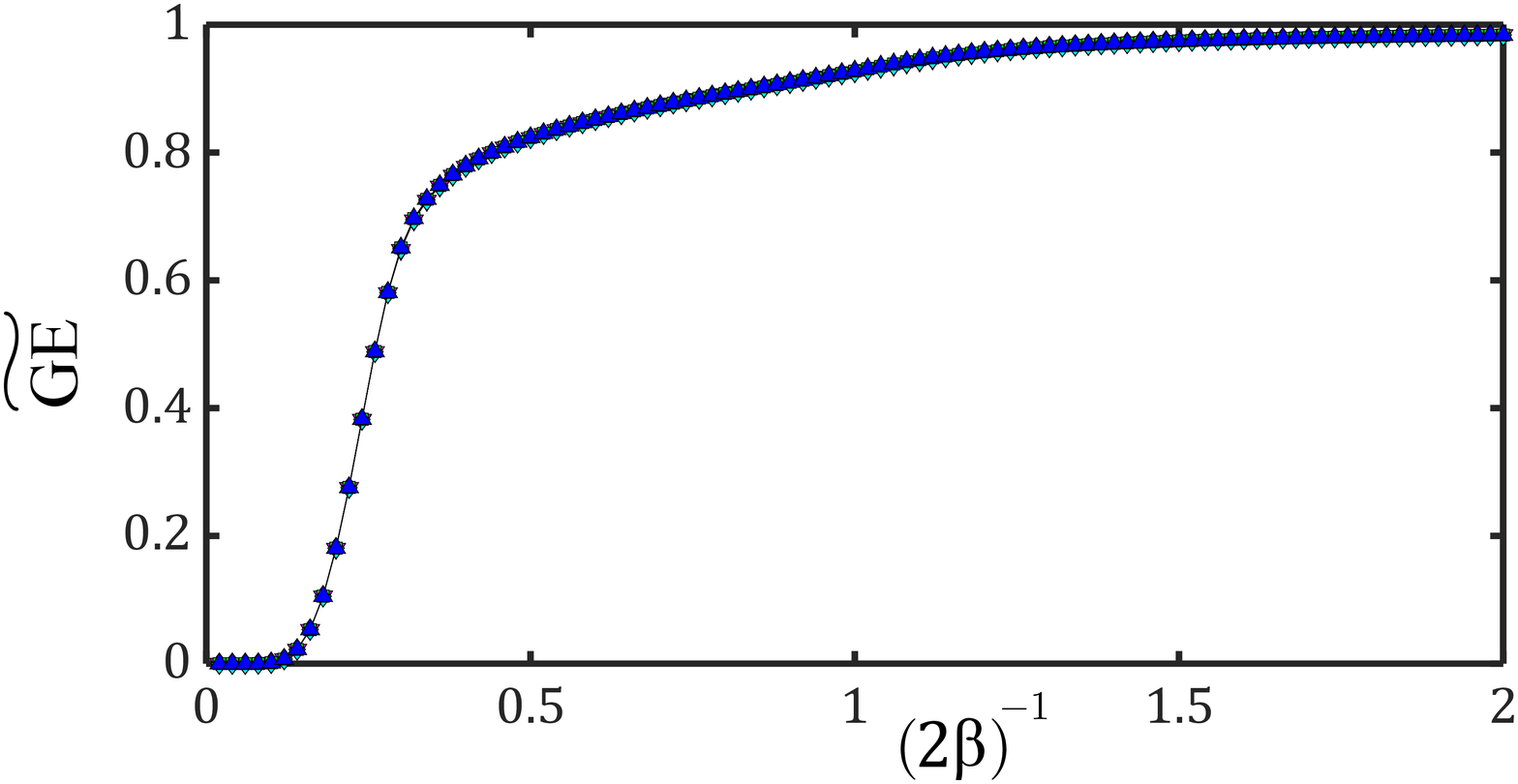}}
	\subfigure[]{\label{fig:3d}\includegraphics[width=60mm, height=40mm]{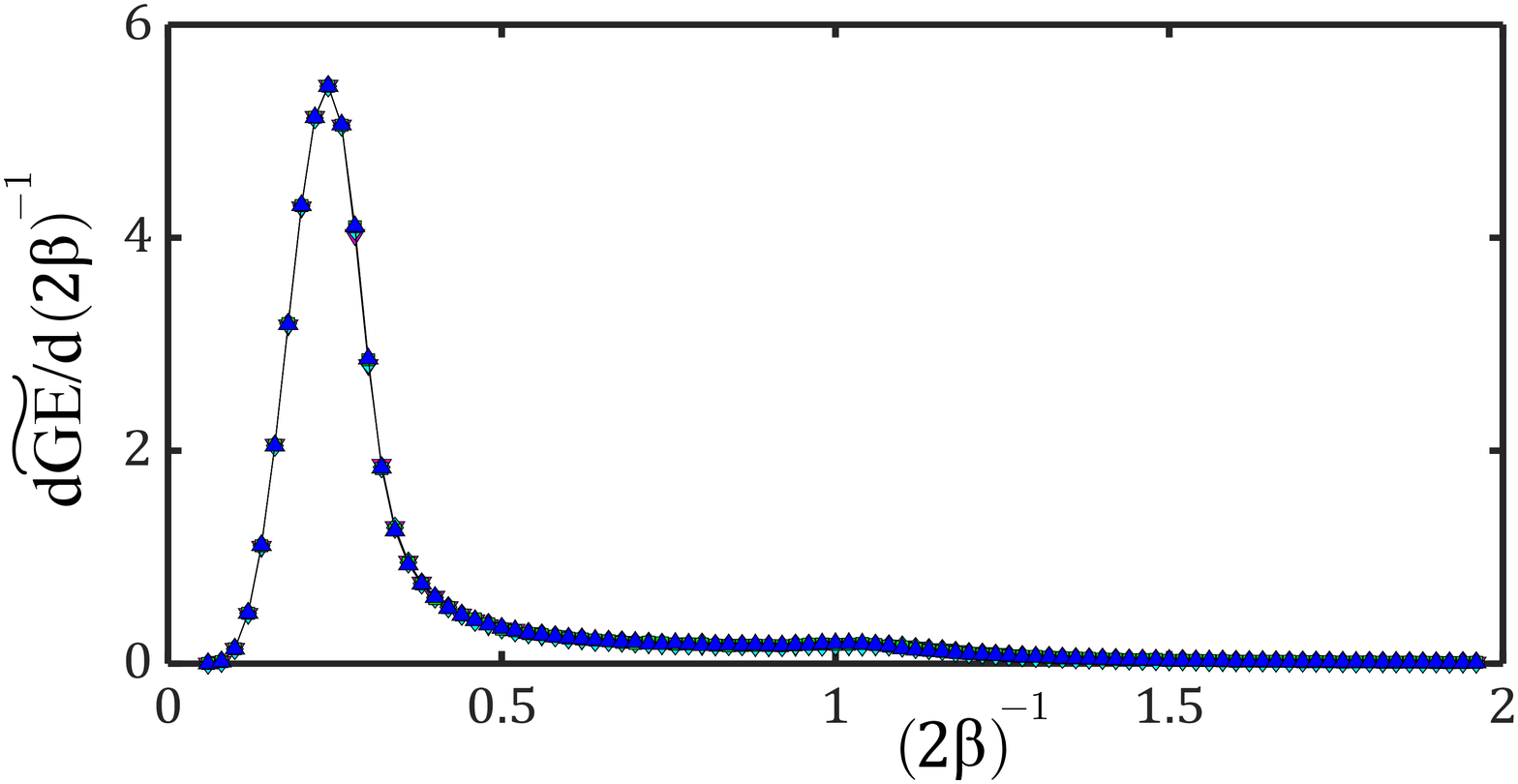}}
	\caption{(Color online) $d=9$ (a) GE (global entanglement) versus $2\beta^{-1}$ (control parameter), (b)
		The first-order derivative of GE. The maximum of $\text{d}\text{GE}/\text{d}(2\beta)^{-1}$ occurs at $0.26$. (c) $\widetilde{\text{GE}}$ (generalized global entanglement) vs $(2\beta)^{-1}$.  (d) The first-order derivative of $\widetilde{\text{GE}}$. The derivative of $\widetilde{\text{GE}}$ peaks at $0.24$. It has been shown that the first critical point occurs at $T^{1}_{c}\approx 0.34$ \cite{li2020accurate}. System sizes are indicated in (a) and no significant finite-size effect is obtained. }\label{fig:3}
\end{figure*}

We can write the partition function in the following form:
\begin{equation}\label{Zclock2}
\mathcal{Z}_{clock}(T)=\sum_{\{n_{i}\}}\text{exp}\{\sum_{\la ij \ra}\frac{\omega ^{n_{i}-n_{j}}+\omega^{n_{j}-n_{i}}}{2T}\},
\end{equation}
where we use the fact $\cos(x)=\frac{e^{ix}+e^{-ix}}{2}$. $i$ and $j$ in Eq. (\ref{Zclock2}) refer to the d-state classical spins placed on the vertices.

On the other hand, in Eq. \ref{ui} we can rewrite the square of the denominator in the form of:
\begin{equation}\label{dnorm}
\begin{aligned}
\sum_{\{m_{i}\}}\la K_{d}|m_{1}m_{2}...m_{N}\ra& \text{exp}\left\{\sum_{i}\beta (\omega^{m_{i}}+\omega^{-m_{i}}) \right\}
\\
&\la m_{1}m_{2}...m_{N}|K_{d}\ra,
\end{aligned}
\end{equation}
where we open a complete basis of eigenvectors of $\hat{Z}$ operator denoted by $|m_i\rangle$ corresponding to each qudit at the edge $i$. By recalling that $|m_{i}\ra$ refers to the state of the qudit at the edge $i$, we can deduce that it corresponds to the edge variable $m_{e}$ in the classical model. Therefore, Eq. (\ref{dnorm}) is nothing more than sum of Boltzmann weights of configurations in the d-state clock model. Indeed, it is the partition function of the clock model up to a constant factor where the control parameter $\beta$ maps to $\frac{1}{2T}$.

Next, using the quantum-classical mapping, it has been shown that the fidelity susceptibility of $|K_d(\beta)\ra$ maps to the heat capacity of the clock model \cite{zarei2020kosterlitz}. Whereas this implies that $|K_d(\beta)\ra$ passes through KT phase transitions for $d\geq 5$ at transition points $\beta^{*}_{1}=\frac{1}{2T_{c}^{1}}$ and $\beta^{*}_{2}=\frac{1}{2T_{c}^{2}}$, it fails to show the critical points $\beta^{*}_{1}$ and $\beta^{*}_{2}$. Indeed, the fidelity susceptibility for $|K_d(\beta)\ra$ similar to the heat capacity for the clock model does not show  a detectable singularity at transition points when $d\geq 5$. Therefore, it is relevant to ask what quantities are reasonable for characterizing critical points in the KT QPTs? In particular, regarding the well-known role of multipartite entanglement in studying different types of quantum phase transitions, we expect that various measures of multipartite entanglement should be useful in the KT QPTs. In the following sections, we consider the above important issue by studying GE as a known measure of multipartite entanglement.
\section{Global Entanglement}\label{sec3}
In this section, we consider GE and $\widetilde{\text{GE}}$ as measures of multipartite entanglement to study QPTs occurring in $|K_d(\beta)\ra$. $\text{GE}$ and $\widetilde{\text{GE}}$ defined by:
\begin{equation}\label{GE}
\text{GE}=\frac{d}{d-1}\left(1-\frac{1}{N}\sum_{i=1}^{N}Tr(\hat\rho_{i}^2)\right),
\end{equation}
and
\begin{equation}\label{tildeGE}
\widetilde{\text{GE}}=\frac{d^2}{d^2-1}\left(1-\frac{2}{N(N-1)}\sum_{(i,j)}Tr(\hat\rho_{i,j}^2)\right),
\end{equation}

where $\hat\rho_{i}$ ($\hat\rho_{i,j}$) is the one- (two-) qudit reduced density matrix corresponding to qudit(s) $i$ $(i,j)$. Note that it is possible to interpret GE ($\widetilde{\text{GE}}$) as average linear entropy of one (two) qudit(s), respectively \cite{rigolin2006operational} where $\text{Tr}(\hat\rho_{i}^{2})$ and $\text{Tr}(\hat\rho_{i,j}^{2})$ are the purity of the quantum state $\hat\rho_{i}$ and $\hat\rho_{i,j}$.

In this regard, we consider two particular qudits denoted by $a,b$ and find reduced density matrices $\hat\rho_{a}$ and $\hat\rho_{a,b}$ which are obtained by taking the partial trace of $|K_d(\beta)\ra\la K_d(\beta)|$. To this end, we write $\hat\rho_{a}$ and $\hat\rho_{a,b}$ in the following form:
\begin{equation}\label{rhoa}
\begin{aligned}
\hat\rho_{a}=\frac{1}{\mathcal{Z}}&\sum_{\{\alpha_{l}=0,1,...,d-1|l\ne a\}}\\
&\la\alpha_{1},\alpha_{2},...,\alpha_{N}|e^{\frac{\beta}{2}\sum_{i}[\hat Z_{i}+\hat Z_{i}^{-1}]}|K_{d}\ra 
\\
&\la K_{d}|e^{\frac{\beta}{2}\sum_{i}[\hat Z_{i}+\hat Z_{i}^{-1}]}|\alpha_{1},\alpha_{2},...,\alpha_{N}\ra .
\end{aligned}
\end{equation}

\begin{equation}\label{rhoab}
\begin{aligned}
\hat\rho_{a,b}=\frac{1}{\mathcal{Z}}&\sum_{\{\alpha_{l}=0,1,...,d-1|l\ne a,b\}}\\
&\la\alpha_{1},\alpha_{2},...,\alpha_{N}|e^{\frac{\beta}{2}\sum_{i}[\hat Z_{i}+\hat Z_{i}^{-1}]}|K_{d}\ra 
\\
&\la K_{d}|e^{\frac{\beta}{2}\sum_{i}[\hat Z_{i}+\hat Z_{i}^{-1}]}|\alpha_{1},\alpha_{2},...,\alpha_{N}\ra .
\end{aligned}
\end{equation}

Interestingly, $\hat\rho_{a}$ and $\hat\rho_{a,b}$ are diagonal matrices, because it is impossible to have a closed loop which acts solely on one or two qudit(s). Therefore, $\hat\rho_{a}$ and $\hat\rho_{a,b}$ are simplified to:
\begin{equation}\label{rhoasimple}
\hat\rho_{a}=\frac{1}{\mathcal{Z}}\left(\mathcal{Z}^a_{0}|0\ra\la 0|+\mathcal{Z}^a_{1}|1\ra\la 1|+...+\mathcal{Z}^a_{d-1}|d-1\ra\la d-1|\right),
\end{equation}

\begin{equation}\label{rhoabsimple}
\begin{aligned}
\hat\rho_{a,b}=&\frac{1}{\mathcal{Z}}(\mathcal{Z}^{ab}_{00}|00\ra\la 00|+\mathcal{Z}^{ab}_{01}|01\ra\la 01|+...
\\
&...+\mathcal{Z}^{ab}_{(d-1)(d-1)}|(d-1)(d-1)\ra\la (d-1)(d-1)|),
\end{aligned}
\end{equation}
\begin{figure*}[t!]
	\centering     %%% not \center
	\subfigure[]{\label{fig:4a}\includegraphics[width=55mm, height=40mm]{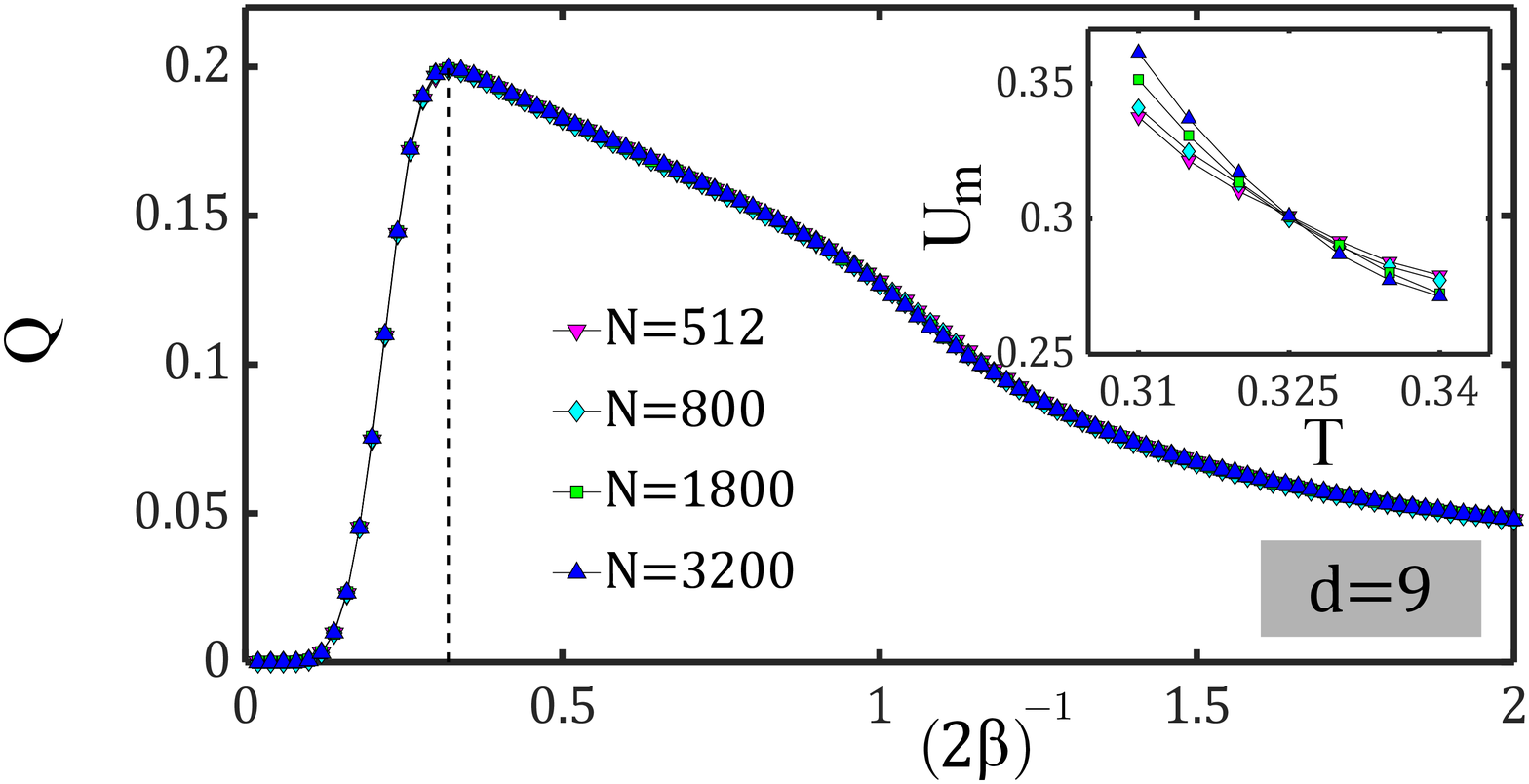}}
	\subfigure[]{\label{fig:4b}\includegraphics[width=55mm, height=40mm]{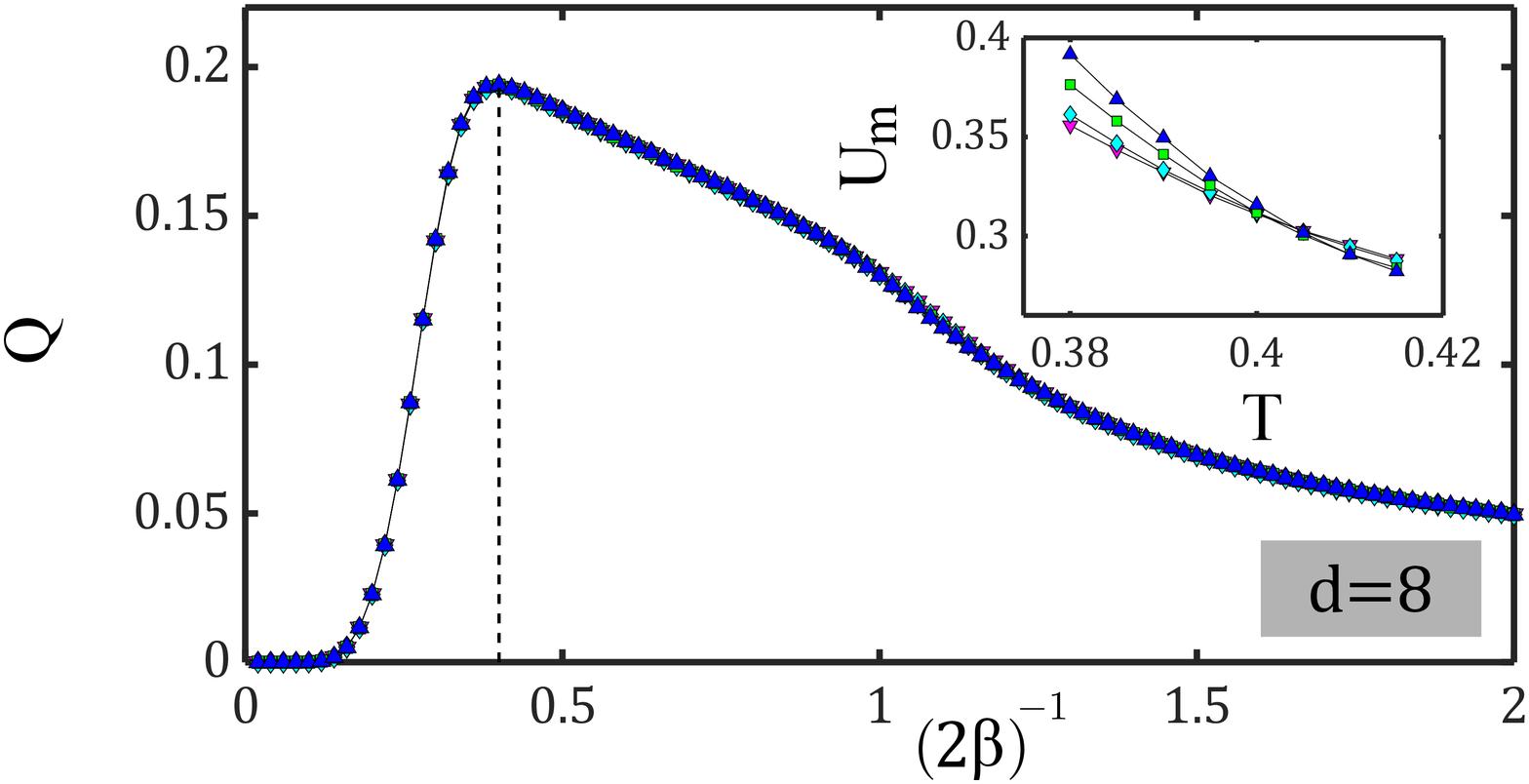}}
	\subfigure[]{\label{fig:4c}\includegraphics[width=55mm, height=40mm]{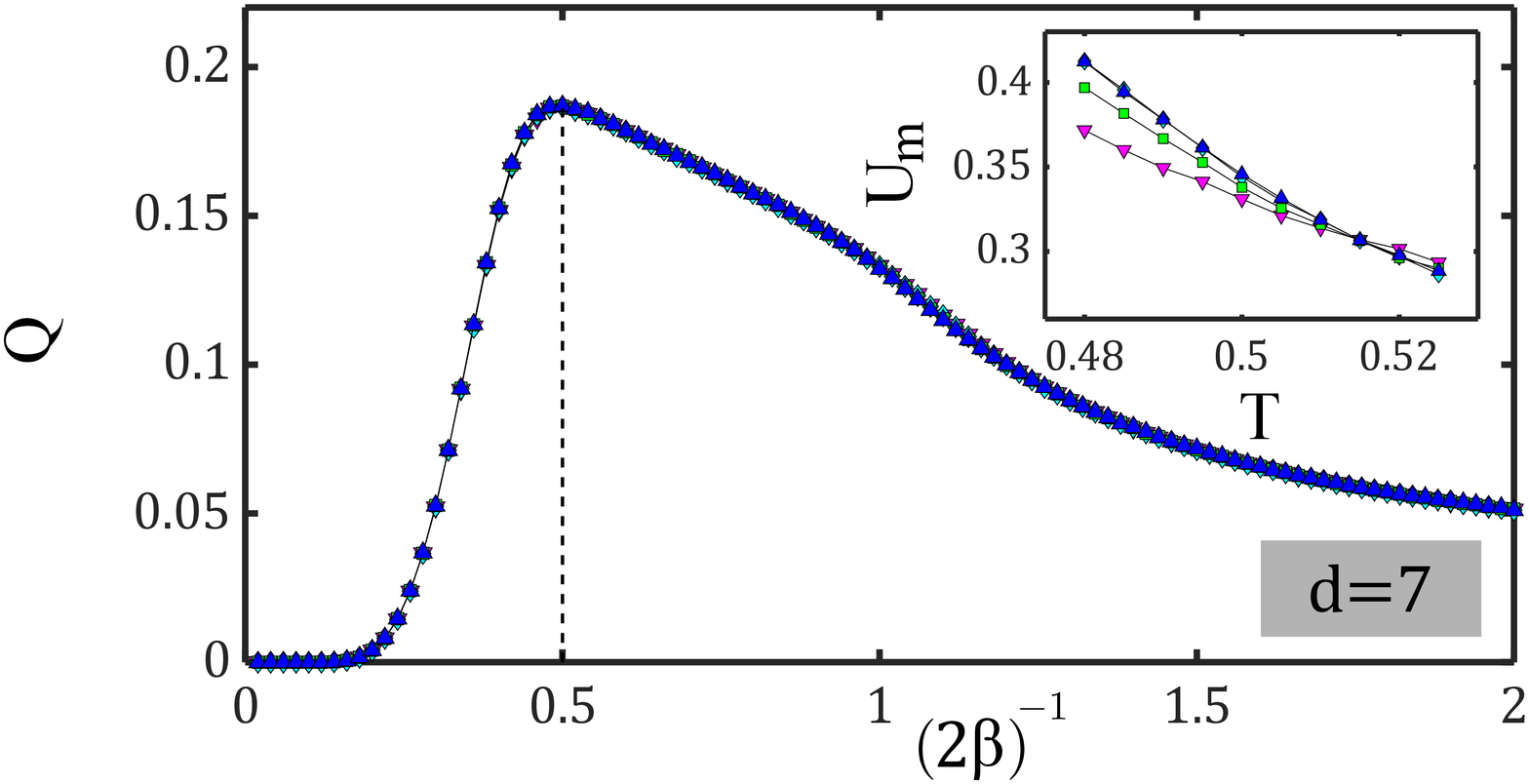}}
	\subfigure[]{\label{fig:4d}\includegraphics[width=55mm, height=40mm]{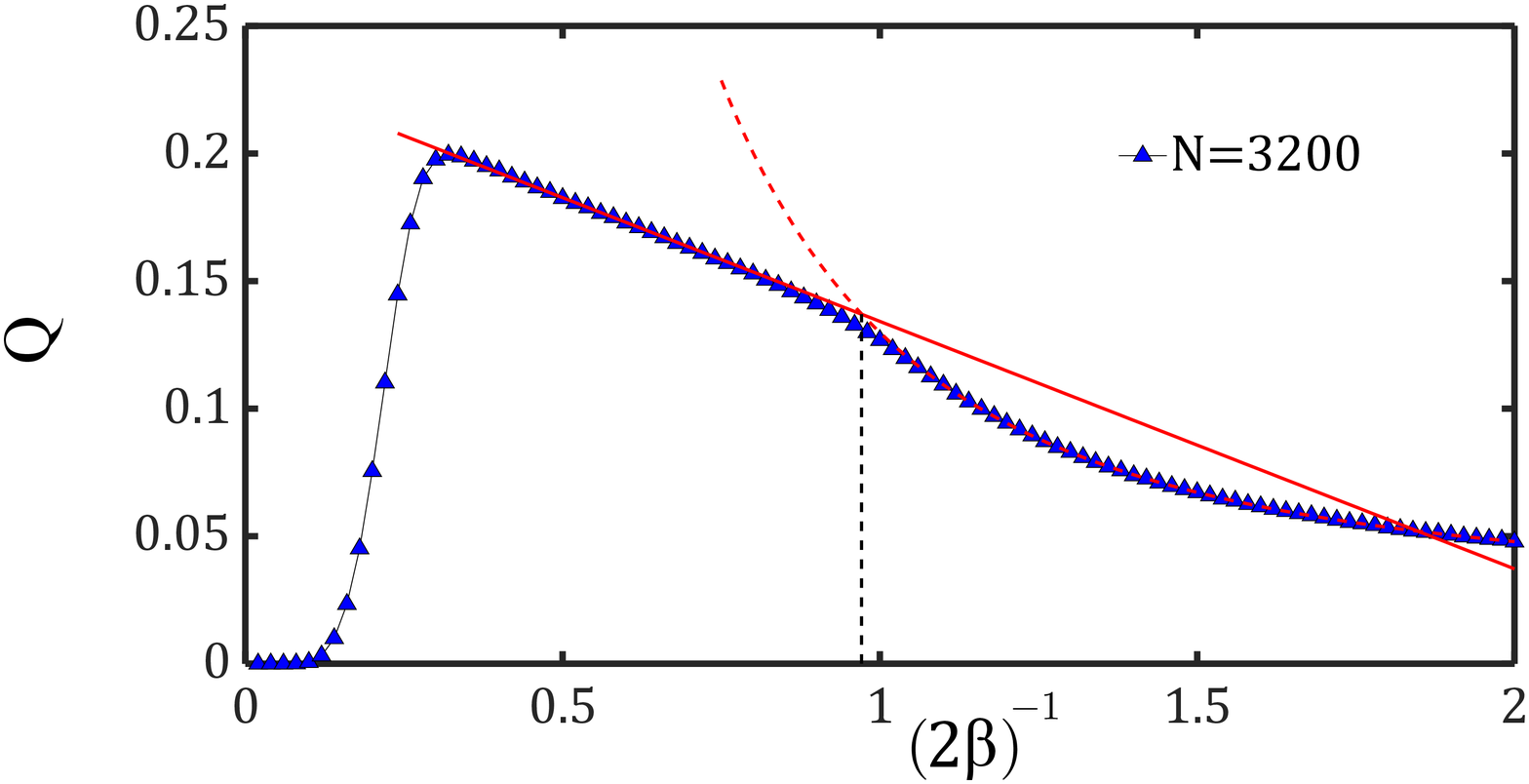}}
	\subfigure[]{\label{fig:4e}\includegraphics[width=55mm, height=40mm]{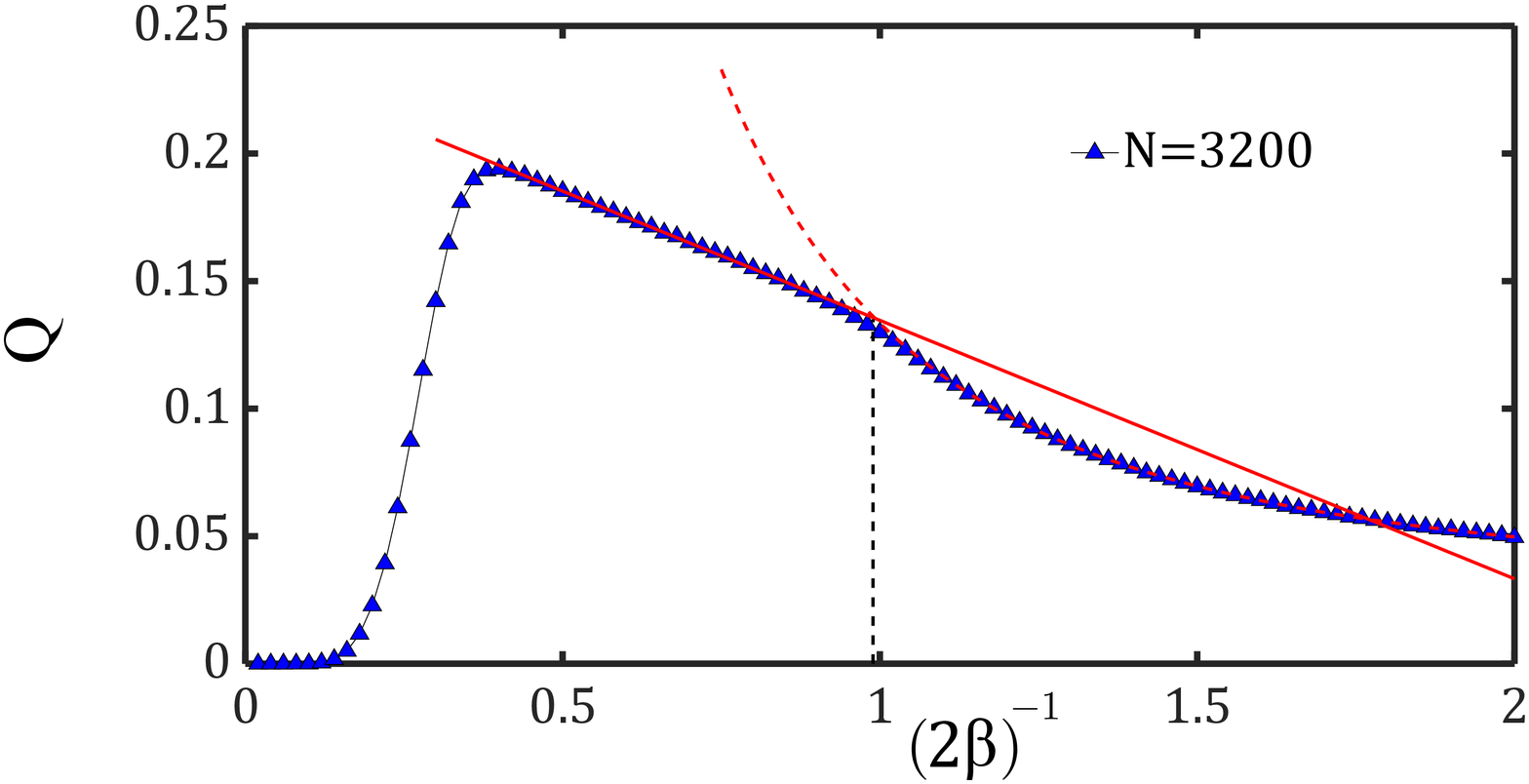}}
	\subfigure[]{\label{fig:4f}\includegraphics[width=55mm, height=40mm]{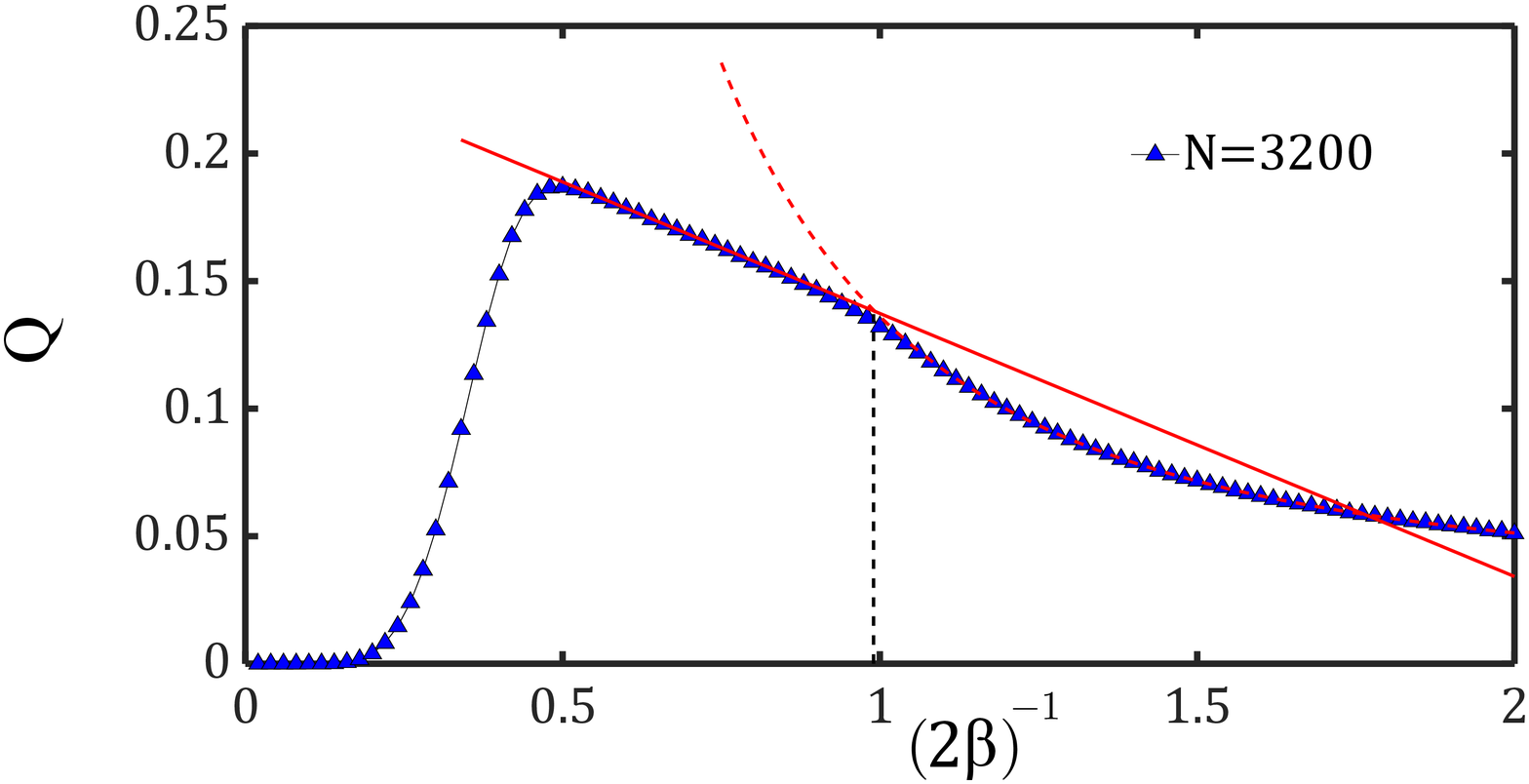}}
	\caption{(Color online) $Q$ (Conditional global entanglement) versus $(2\beta)^{-1}$ (control parameter) for (a) $d=9$, (b) $d=8$ and (c) $d=7$.  The insets show $U_{m}$ (cumulant) versus T (temperature) in the clock model. Q and the fitted functions for (d) $d=9$ , (e) $d=8$ and (f) $d=7$ when $N=3200$. For $d=9$ the fitted functions are: $y=-0.097x+0.2312$ and $y=0.1018x^{-2.358}+0.02804$. For $d=8$ the fitted functions are:
		$y=-0.1013x+0.2359$ and $y=0.1047x^{-2.321}+0.02879$. For $d=7$ the fitted functions are: $y=-0.1031x+0.2404$ and $y=0.1067x^{-2.289}+0.02943$. System sizes are indicated in (a) and no significant finite-size effect is obtained.}  \label{fig:4}
\end{figure*} \\ 
where $\mathcal{Z}^{a}_{m}/\mathcal{Z}$ $(\mathcal{Z}^{ab}_{mm^{\prime}}/\mathcal{Z})$ is the sum of the squares of probability amplitudes of configurations in which the state of qudit(s) $a$ $(a,b)$ is  $|m\ra$ $(|mm^{\prime}\ra)$. According to the quantum-classical mapping described above, we can construct a relation between $\mathcal{Z}^{a}_{m}/\mathcal{Z}$ $(\mathcal{Z}^{ab}_{mm^{\prime}}/\mathcal{Z})$ and classical quantities in the clock model. To this end, remind that corresponding to neighboring classical spins we defined an edge variable of $m$ in the clock model. In this regard, similar to the definition of the partition function in Eq. \ref{Zclock2}, $\mathcal{Z}^{a}_{m}$ $(\mathcal{Z}^{ab}_{mm^{\prime}})$ is also the sum of Boltzmann weights of spin configurations of the classical 
clock model, where the absolute value of the edge variable(s) corresponding to edge(s) $a$ $(a,b)$ is (are) fixed to $m$ $(m,m^{\prime})$. Therefore, we can finally write GE and $\widetilde{\text{GE}}$ in the following way:
\begin{equation}\label{sGE}
\text{\text{GE}}=\frac{d}{d-1}\left(1-\frac{1}{N}\sum_{a=1}^{N}[(P^{a}_{0})^{2}+(P^{a}_{1})^{2}+...+(P^{a}_{d-1})^{2}]\right),
\end{equation}
and
$$\widetilde{\text{GE}}=\frac{d^2}{d^2-1}$$
\begin{equation}\label{stildeGE}
\left(1-\frac{2}{N(N-1)}\sum_{(a,b)}[(P^{ab}_{00})^2+...+(P^{ab}_{(d-1)(d-1)})^{2}]\right),
\end{equation}
where $P^{a}_{m}=\mathcal{Z}^{a}_{m}/\mathcal{Z}$ $(P^{ab}_{mm^{\prime}}=\mathcal{Z}^{ab}_{mm^{\prime}}/\mathcal{Z})$ is equal to the probability that the classical system in the temperature of $T$ chooses spin configurations wherein the absolute of edge variable(s) of a particular link(s) $a$ $(a,b)$ has (have) the value(s) of $m$ $(m,m^{\prime})$.

By the above mapping, we are able to calculate GE and $\widetilde{\text{GE}}$ for our quantum model to see if they are capable of characterizing the QPTs induced in $|K_d(\beta)\ra$ . To achieve this aim, we simulate the d-state clock model from $d=7$ to $d=9$ for different system sizes up to $N=3200$. We find that while both quantities behave monotonically, the derivatives of $\text{GE}$ and $\widetilde{\text{GE}}$ do not peak at or near the critical points. For $d=9$, the maximum of the derivative of GE occurs at $0.26$ while the derivative of $\widetilde{\text{GE}}$ peaks at $0.24$. This is while the first critical point has been obtained at $\text{T}^{1}_{c}\approx 0.34$ \cite{li2020accurate}. Figure \ref{fig:3} shows the behavior of $\text{GE}$, $\widetilde{\text{GE}}$, and their derivatives for $d=9$. Similar figures are also obtained for $d=8$ and $d=7$. In the case of $d=8$, the peak position of $\text{dGE/d}(2\beta)^{-1}$ is at $0.32$ and the maximum of $\text{d}\widetilde{\text{GE}}/\text{d}(2\beta)^{-1}$ occurs at $0.30$. In comparison, the first transition point is at $\text{T}^{1}_{c} \approx 0.42$ \cite{li2020accurate}. When $d=7$, $\text{dGE/d}(2\beta)^{-1}$ becomes maximum at $0.42$ and $\text{d}\widetilde{\text{GE}}/\text{d}(2\beta)^{-1}$ peaks at $0.38$, whereas it has been shown that the first critical point is at $\text{T}^{1}_{c} \approx 0.54$ \cite{li2020accurate}. These results suggest that both GE and $\widetilde{\text{GE}}$ are not able to detect KT QPTs. 

Regarding the above results for $\text{GE}$ and $\widetilde{\text{GE}}$, it becomes an important task to consider which other measure of multipartite entanglement is able to capture a KT QPT. In particular, in \cite{samimi2022global} it has been shown that a conditional version of global entanglement Q, which is equal to the difference between $\widetilde{\text{GE}}$ and $\text{GE}$, is a good measure for characterizing TQPTs. In this regard, we study this quantity for the KT QPT. 

\section{QUANTUM CONDITIONAL ENTANGLEMENT}\label{sec4}
As seen in the previous section, both $\text{GE}$ and $\widetilde{\text{GE}}$ show monotonic behavior versus the control parameter $(2\beta)^{-1}$. Their derivatives do not also peak near the critical points and hence they fail to quantify the transition points in the KT QPT. On the other hand, as it has been shown in \cite{samimi2022global} it seems that it is the presence of long-range entanglement in the topological phase which leads to the monotonic behavior of GE. However, the quantity $Q=\widetilde{\text{GE}}-\text{GE}$ is able to cancel out the effect of long-range entanglement and peaks at the critical point in the case of ordinary TQPT. In this regard, we intend to study the behavior of $Q$ for $|K_d(\beta)\ra$ and ask if it is still a reasonable indicator of KT QPT. 

To this end, we use the classical-quantum mapping and calculate $Q$ by simulating the d-state clock model. We consider $Q$ in terms of $(2\beta)^{-1}$ for $7\leq d\leq 9$ up to system sizes $N=3200$ . Interestingly, as it is shown in Fig. \ref{fig:4}, $Q$ develops a peak near the first critical point. However, the peak position of Q, denoted hereafter by $T_{Q}$, is slightly different from the transition point which is obtained from very recent large scale simulation of clock model \cite{li2020accurate,chen2022monte,li2020critical}. To illustrate the effect of small system size, we obtain transition points by evaluating a cumulant $U_m$ \cite{chatterjee2018ordering} with the use of system sizes up to $N=3200$ and compare the results with $T_{Q}$. The parameter $U_m$ is:
\begin{equation}\label{Um}
U_{m}=1-\frac{\la m^{4}_{\phi}\ra}{2\la m^{2}_{\phi}\ra^{2}},
\end{equation}
where $\phi=\text{tan}^{-1}(\sigma_{y}/\sigma_{x})$, $\sigma_{x}=\sum^{N/2}_{i=1}\cos\theta _{i}$, $\sigma_{y}=\sum^{N/2}_{i=1}\sin\theta _{i}$ and $m_{\phi}=\la \cos(d\phi)\ra$ is the effective order parameter. Then, the transition point is characterized from the intersection of $U_m$ for various $N$. The inset in Fig. \ref{fig:4} shows $U_{m}$ versus T for $7\leq d\leq 9$.

 Table \ref{table:1} compares $T_{Q}$ and the intersection position of $U_m$ which is denoted by $T_{U_{m}}$. \begingroup 
 \begin{table}[h!]
 	\setlength{\tabcolsep}{10pt} % Default value: 6pt
 	\renewcommand{\arraystretch}{2} % Default value: 1
 	\begin{tabular}{|c|c| c|}
 		\hline
 		d&{$T_{U_m}$}&{$T_{Q}$}\\
 		\hline
 		\hline
 		% after \\: \hline or \cline{col1-col2} \cline{col3-col4} ...
 		7&0.513(2)&0.492(7)\\
 		\hline
 		8&0.405(2)&0.393(3)\\
 		\hline
 		9&0.326(2)&0.322(5)\\
 		\hline
 	\end{tabular}
 	\caption{The second column corresponds to the transition points which are obtained from the intersection of $U_{m}$ (cumulant). Third column shows $T_{Q}$ (the peak position of Q).}\label{table:1}
 \end{table}
 \endgroup
 Despite our limited system sizes, the results show good consistency in characterizing the first transition points.

Now let us move to the second critical point. As can be seen from Fig. \ref{fig:4}, the behavior of Q within the KT phase differs from the topological phase. One can readily fit a linear function to Q in the KT phase. In comparison, a power-law curve fits well to the data in the topological phase. In order to capture the starting position of the change of behavior of Q, we plot the fitted linear and the power-law function and consider their intersection point. In this approximation, which is clearly vulnerable to numerical uncertainty, we obtain  $(2\beta)^{-1}\approx 0.97~~ (d=9)$, $(2\beta)^{-1}\approx 0.99~~ (d=8)$ and $(2\beta)^{-1}\approx 0.99~~ (d=7)$. In previous studies with large system sizes, the second transition points for $d=9,8$ and $d=7$ are reported as $T^{2}_{c} \approx 0.91$ \cite{li2020critical,li2020accurate}.  Therefore, Q seems to capture the second transition point as well. 

 On the other hand, it is known that the behavior of a suitable measure of multipartite entanglement should be affected by the divergence of the correlation length at the critical point. It particularly leads to the expectation that a multipartite measure should be maximum at the critical point. However, for KT phase transitions the problem is more interesting because the correlation length diverges in the entire KT phase. In particular, in \cite{li2020critical} it has been shown that the correlation length in the clock model in the finite system sizes is maximum at the first transition point and it decreases monotonically with the increase of temperature. This kind of behavior is similar to the behavior of Q within the KT phase in our model. Consequently, it suggests that Q is a suitable measure for studying multipartite entanglement in QPTs.

\section{concluding remarks}\label{sec5}
It has been shown that GE and $\widetilde{\text{GE}}$ are able to detect the transition point in the TQPT \cite{samimi2022global}. Therefore, one might expect that they can also serve as reasonable tools for detecting KT QPTs. However, our work shows that they are not sensitive at transition points of KT QPTs. 
Despite GE and $\widetilde{\text{GE}}$, $Q$ 
works relatively well for not only detecting transition points in KT QPTs, but also for distinguishing different phases by showing noticeably different behavior in each phase. Particularly, the monotonic decrease of $Q$ within the KT phase is generally similar to the behavior of the correlation length in the clock model \cite{li2020critical}. Hence, it is possible to regard Q as a suitable measure of multipartite entanglement to characterize QPTs \cite{samimi2022global}.

We also note that we studied the behavior of Q, GE and 
$\widetilde{\text{GE}}$ for $\text{d}=5,6$ wherein the system undergoes KT QPTs. The results are similar to those given by $\text{d}=7,8,9$. 
However, there exists more discrepancy between $T_{U_m}$ 
and $T_Q$ in these cases. Considering the fact that KT QPTs for $\text{d}=5,6$ are somewhat controversial and challenging to consider numerically \cite{lapilli2006universality,hwang2009six,baek2010comment,baek2010non,baek2013residual,kumano2013response,chatelain2014dmrg,surungan2019berezinskii}, such discrepancy might be expected for finite system sizes considered in our study. However, we expect that more detailed and computationally expensive studies in the future can elucidate the importance of Q in various QPTs.

\section*{Acknowledgement}
We would like to thank A. Ramezanpour for fruitful discussions.
%apsrev4-2.bst 2019-01-14 (MD) hand-edited version of apsrev4-1.bst
%Control: key (0)
%Control: author (8) initials jnrlst
%Control: editor formatted (1) identically to author
%Control: production of article title (0) allowed
%Control: page (0) single
%Control: year (1) truncated
%Control: production of eprint (0) enabled

%\section*{References}\label{section.rr}
%\bibliography{ref}
%apsrev4-2.bst 2019-01-14 (MD) hand-edited version of apsrev4-1.bst
%Control: key (0)
%Control: author (8) initials jnrlst
%Control: editor formatted (1) identically to author
%Control: production of article title (0) allowed
%Control: page (0) single
%Control: year (1) truncated
%Control: production of eprint (0) enabled
%

\end{document}